# Physical properties of niobium based intermetallics (Nb$_3$B; $B$ = Os, Pt, Au): a DFT based *ab-initio* study

M. I. Naher[a], F. Parvin[a], A. K. M. A. Islam[b], S. H. Naqib[a*]

[a]Department of Physics, University of Rajshahi, Rajshahi 6205, Bangladesh
[b]International Islamic University Chittagong, 154/A College Road, Chittagong-4203, Bangladesh

*Corresponding author: salehnaqib@yahoo.com

## Abstract

Structural, elastic and electronic band structure properties of A-15 type Nb-based intermetallic compounds Nb$_3$B ($B$ = Os, Pt, Au) have been revisited using first principles calculations based on the density functional theory (DFT). All these show excellent agreement with previous reports. More importantly, electronic bonding, charge density distribution and Fermi surface features have been studied in detail for the first time. Vickers hardness of these compounds is also studied. The Fermi surfaces of Nb$_3$B contain both hole- and electron-like sheets, the features of which change systematically as one move from Os to Au. The electronic charge density distribution implies that Nb$_3$Os, Nb$_3$Pt and Nb$_3$Au have a mixture of ionic and covalent bondings with a substantial metallic contribution. The charge transfer between the atomic species in these compounds has been explained via the Mulliken bond population analysis and the Hirshfeld population analysis. The bonding properties show a good correspondence to the electronic band structure derived electronic density of states (DOS) near the Fermi level. Debye temperature of Nb$_3$B ($B$ = Os, Pt, Au) have been estimated from the elastic constants and show a systematic behavior as a function of the $B$ atomic species. We have discussed implications of the results obtained in this study in details in this paper.

**Keywords:** Intermetallic compounds; Density functional theory; Elastic constants; Electronic band structure; Fermi surface; Bonding characteristics

## 1. Introduction

In the progress of many technologically important science and engineering sectors, intermetallic compounds have played a notable role. Intermetallic compounds are one of the oldest and most important condensed phases, being a subject of constant interest of physicists, chemists and materials scientists [1-8]. Large number of intermetallic compounds exhibit attractive combination of physical and mechanical properties, including high melting point, low density and good oxidation or corrosion resistance. They have vast applications in aerospace industry, aircraft and automotive engine, biomedical instrumentations, microelectronics, electronic devices, batteries, hydrogen storage systems, and chemical industries. In every solder joint in an electronic circuit, between the solder and the system of interest, a layer is present containing one or more intermetallic compounds [9-11].

A great deal of effort has been made in the last few decades to raise the critical temperature $T_c$ of conventional superconducting materials. It is well established that the compounds with A-15 type structure sometimes exhibit superconductivity. In this context the A-15 family of $A_3B$ compounds have also been studied extensively both theoretically and experimentally [12-28]. Even in normal state, the studies of the physical properties of these materials are scientifically informative and interesting [29-31]. Intermetallic compounds $Nb_3B$ ($B$ = Os, Pt, Au) shows superconductivity with widely varying transition temperatures [28]. X-ray photoemission spectra of $Nb_3B$ ($B$ = Os, Pt, Au) have shown that the energy bands of Nb 4$d$ and 5$d$ orbital of these $A_3B$ systems seem to be more and more separated with increasing atomic number of the $B$ element [32–35]. For these reasons, physical properties of $Nb_3B$ compounds such as structural, elastic, electronic, bonding, optical, thermal, etc., are quite instructive to know. To the best of our knowledge, the elastic properties (except bulk modulus), detailed electronic properties (band structure, density of states, Fermi surface features), charge density mapping, Mulliken bond population and hardness of $Nb_3B$ ($B$ = Os, Pt, Au) compounds have not been studied yet. Therefore, we undertook this project to explore some of these properties in this study.

We are interested to revisit the structural and mechanical properties because they are important for different application related physical parameters. Comparison of our results with the previous studies will lend strong support to the methodology employed in this paper. These properties are also related to different fundamental solid state and thermal properties. The electronic properties, such as electronic band structure, density of states, Fermi surface and charge density are related to charge transport, optical and electronic thermal processes. Mulliken bond population analysis elucidates the bonding nature of the compounds under study. All the mechanical properties are related to the bonding characteristics.

This paper is organized as follows. In Section 2, computational scheme is described in brief. Section 3 consists of results and analysis of all the physical properties under study. Finally, in Section 4 we have discussed the theoretical results in detail and have drawn the main conclusions of this study.

## 2. Computational Scheme

CASTEP (Cambridge Serial Total Energy Package) code [36] has been used extensively to explore various properties of materials belonging to different classes. This code is based on the density functional theory (DFT), which employs a total energy plane wave pseudopotential method [37, 38]. In this investigation, the interaction between the valence electrons and ion cores has been represented by the Vanderbilt-type ultra-soft pseudopotential for Nb, Os, Pt and Au atoms [39]. The exchange-correlation energy is calculated using the Generalized Gradient Approximation (GGA) of the Perdew–Burke–Ernzerhof (PBE) scheme [40]. The valence electron configurations used in this research were $4s^2\ 4p^6\ 4d^4\ 5s^1$ for Nb, $5s^2\ 5p^6\ 5d^6\ 6s^2$ for Os, $5s^2\ 5p^6\ 5d^9$, $6s^1$ for Pt and $5s^2\ 5p^6\ 5d^{10}$, $6s^1$ for Au, respectively. The Brillouin zone (BZ) integrations have been performed using the Monkhorst and Pack [41] *k*-point meshes. For geometry optimization 16×16×16 *k*-mesh has been used to integrate the wave function in the BZ. A cut-off energy of 500 eV has been used. The Fermi surfaces were obtained by sampling the whole BZ with the *k*-point meshes 25×25×25. These parameters are sufficient for well converged total energy, geometrical configurations and elastic moduli. The geometry optimization of $Nb_3B$ was done using the Broyden–Fletcher–Goldfarb–Shanno (BFGS) minimization scheme [42]. Optimization is performed using convergence thresholds of $10^{-5}$ eV/atom for the total energy and $10^{-3}$ Å for maximum lattice point displacement. Maximum stress and force were 0.05 GPa and 0.03 eV/Å, respectively for all calculations. The elastic constants were determined by applying a set of homogenous deformations with a finite value and calculating the resulting stresses [43]. The elastic stiffness coefficients have been determined from a linear fit of the calculated stress as a function of theoretical strain.

After optimization, the elastic constants, electronic band structure, energy density of states, charge density distribution, Mulliken population were calculated. From the bonding analysis Vickers hardness of the compounds were estimated. For all the equilibrium structures, the Mulliken populations were investigated using a projection of the plane-wave states onto a linear combination of atomic orbital basis sets [44-45], which is widely used to perform charge transfers and bond population analysis.

## 3. Physical Properties of $Nb_3B$

### *3.1. Structural Properties*

Like other niobium based A-15 intermetallic compounds, $Nb_3B$ assumes cubic structure with space group *Pm-3n* (no. 223), consisting of six *A* atoms that lie on the surface and form chains along the axis directions. There are two *B* atoms per unit cell which occupy the bcc sites in the unit cell. The Nb and *B* atoms occupy the following Wyckoff positions in the unit cell [46-47], Nb atoms: (0.25, 0, 0.5); (0.5, 0.25, 0); (0, 0.5, 0.25); (0.75, 0, 0.5); (0.5, 0.75, 0); (0, 0.5, 0.75) and *B* atoms: (0, 0, 0); (0.5, 0.5, 0.5). Fig.1. shows the crystal structures of these compounds. The unit cell consists of two formula units (Z = 2) and 8 atoms. Table 1 summarized the results of first-principle calculations of the structural properties, together with available experimental and theoretical values [48-50] for comparison. Calculated values of lattice parameters are in excellent agreement with those found in previous studies [48-50]. It is seen that the lattice constant and the cell volume increases slightly as one move from Os to Au. The elastic moduli, on the other hand decrease substantially. This implies that the bonding strength decreases as the atomic number increases in these $Nb_3B$ compounds. The optimized cell parameters of these compounds were also calculated by taking into account of the spin orbit coupling (SOC). The cell parameters are almost identical to those reported in Table 1. Therefore, as far as the structural parameters are concerned, SOC plays no significant role. The effect on electronic band structure is also minimal.

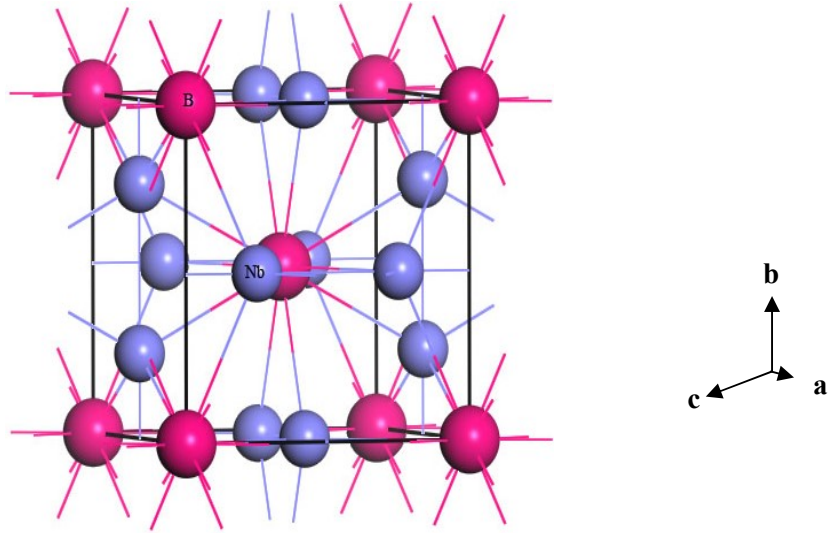

**Figure 1.** Crystal structure of Nb$_3$B (B = Os, Pt, Au) unit cell.

**Table 1**

Calculated and experimental lattice constants $a$ (Å), equilibrium volume $V_o$ (Å$^3$), bulk modulus $B$ (GPa) of Nb$_3$Os, Nb$_3$Pt and Nb$_3$Au.

| Compounds | $a$ | $V_o$ | $B$ | Ref. |
|---|---|---|---|---|
| Nb$_3$Os | 5.16 | 137.20 | 225.26 | This |
| | 5.14 | - | - | [48,49]$^{Exp.}$ |
| | 5.18 | - | 218.78 | [50]$^{Theo.}$ |
| Nb$_3$Pt | 5.19 | 139.74 | 199.96 | This |
| | 5.15 | - | - | [48,49]$^{Exp.}$ |
| | 5.20 | - | 201.76 | [50]$^{Theo.}$ |
| Nb$_3$Au | 5.24 | 144.11 | 182.62 | This |
| | 5.20 | - | - | [48,49]$^{Exp.}$ |
| | 5.24 | - | 180.88 | [50]$^{Theo.}$ |

### *3.2. Mechanical Properties*

Elastic constants of solids provide with a link between the mechanical and dynamical behavior of the crystals and give important information about their response to the external stress as characterized by the bulk modulus $B$, shear modulus $G$, Young's modulus $Y$ and Poisson's ratio, ν. These constants play an important role in determining the strength of the material and are important practical indicators for applications. They also provide with information regarding the mechanical stability of the system. For the material with cubic symmetry, there are only three independent single crystalline elastic constants: $C_{11}$, $C_{12}$, and $C_{44}$. The calculated elastic constants are tabulated in Table 2. The requirement of mechanical stability in a cubic structure leads to the following Born conditions [51]: $C_{11}+2C_{12} > 0$, $C_{44} > 0$, $C_{11}-C_{12} > 0$. All the compounds under study satisfy these criteria. The tetragonal shear modulus, $C'$, which is a measure of stiffness, is related to $C_{11}$ and $C_{12}$ by the following relation: $C' = (C_{11}- C_{12})/2$.

The polycrystalline elastic moduli can be calculated from the single crystalline elastic constants. Calculated values of bulk modulus, $B$ and shear modulus, $G$ are given in Table 3. To get the Young's modulus ($Y$), Poisson's ratio (ν), and shear anisotropy factor ($A$) at zero pressure, following well known identities are used [52]:

$$Y = \frac{9BG}{(3B + G)} \quad (1)$$

$$\nu = \frac{(3B - 2G)}{2(3B + G)} \quad (2)$$

and $$A = \frac{2C_{44}}{(C_{11} - C_{12})} \quad (3)$$

We observe that, $C_{44}$, which reflects the resistance to shear deformation, is lower than $C_{11}$, which is related to the unidirectional compression along the principle crystallographic directions. This means that the cubic cell is more easily deformed by a shear in comparison to the unidirectional compression. Since the elastic moduli and the melting temperatures of solids are both determined by the bonding strength of the material, it is not surprising that these parameters are linked to each other [53-54]. By comparing the elastic moduli of the compounds, one can say that the melting point of $Nb_3Os$ should be the highest and that of $Nb_3Au$, the lowest.

For all three compounds $B > G$ (Table 3), indicating that the shear modulus limits the mechanical stability. The value of Poisson's ratio is linked to the bonding nature of materials and is different, for different dominating bonding [55]. For covalent materials, the value of ν is small (typically ν ~ 0.10); whereas for ionic materials, the typical value of ν is ~ 0.25; for metallic materials, ν is typically ~ 0.33 [55 - 57]. The values of ν for $Nb_3Pt$ and $Nb_3Os$ are found to be 0.31 and 0.32, respectively.

Some of the factors on which the brittle and ductile nature of a given system depends are as follows: Pugh's index ($G/B$), Poisson's ratio (ν) and Cauchy pressure ($C_{12} - C_{44}$). Pugh [58-59] proposed an empirical relationship to distinguish the mechanical properties (ductility and brittleness) of materials. A material should be ductile if the $G/B$ is less than 0.5. Otherwise it is brittle. In our case, for all the compounds, $G/B$ is much less than 0.5, i.e., these compounds should behave in a ductile manner. $Nb_3Au$ is expected to be the most ductile. Poisson's ratio is also an indicator of brittle/ductile behavior. For brittle materials ν ~ 0.10, whereas for ductile materials, ν is typically 0.33 [55]. Here, the Poisson's ratio of $Nb_3Os$, $Nb_3Pt$ and $Nb_3Pt$ are 0.32, 0.31 and 0.33, respectively, which again suggests that the compounds should behave in ductile manner. On the other hand, a positive Cauchy pressure suggests ductility for a material, whereas a negative value suggests brittleness [60]. The calculated values of Cauchy pressure for $Nb_3Os$, $Nb_3Pt$ and $Nb_3Pt$ are given in Table 2. The Cauchy pressures of the compounds are positive, which also indicate that both these compounds are ductile in nature.

The machinability index $\mu_M$, due to Sun *et al.* [61], is defined as,

$$\mu_M = B/C_{44}$$

which may be interpreted as a measure of plasticity [62-64]. Large value of $B/C_{44}$ indicates that the materials under study have excellent lubricating properties. The $B/C_{44}$ values for $Nb_3Os$, $Nb_3Pt$ and $Nb_3Au$ are 3.35, 3.15 and 4.17, respectively. Since $B$ is a weighted average of $C_{11}$ and $C_{12}$ and the condition for mechanical stability requires that $C_{12}$ be smaller than $C_{11}$, we are then left with the result that $B$ is required to be intermediate in value between $C_{11}$ and $C_{12}$: $C_{12} < B < C_{11}$ [65, 66], as observed in this study.

**Table 2**

The calculated elastic constants, $C_{ij}$ (GPa), Cauchy pressure, $(C_{12} - C_{44})$ (GPa) and tetragonal shear modulus, $C'$ (GPa) for $Nb_3B$ at $P = 0$ GPa and $T = 0$ K.

| Compounds | $C_{11}$ | $C_{12}$ | $C_{44}$ | $(C_{12}-C_{44})$ | $C'$ |
|---|---|---|---|---|---|
| $Nb_3Os$ | 413.09 | 131.34 | 67.26 | 64.07 | 140.38 |
| $Nb_3Pt$ | 373.58 | 113.15 | 63.57 | 49.58 | 130.21 |
| $Nb_3Au$ | 334.46 | 106.71 | 47.18 | 59.53 | 113.88 |

Shear anisotropy factor ($A$) is an important parameter related to the structural stability, defect dynamics and elastic anisotropy of crystals [67]. For example a defect tends to propagate along the '*easy direction*' inside the crystal. For an isotropic system, $A = 1$, while any value deviating from unity points towards an elastic anisotropy. The calculated values of $A$ for $Nb_3Os$, $Nb_3Pt$ and $Nb_3Au$ are 0.48, 0.49 and 0.41, respectively, implying significant anisotropy. $Nb_3Au$ is the most anisotropic.

**Table 3**

The calculated bulk modulus $B$ (in GPa), shear modulus $G$ (in GPa), Young's modulus $Y$ (in GPa), Pugh's indicator $G/B$, machinability index $B/C_{44}$, Poisson's ratio $v$, anisotropy factor $A$ for Nb$_3$B at $P = 0$ GPa and $T = 0$ K.

| Compounds | $B$ | $G$ | $Y$ | $G/B$ | $B/C_{44}$ | $v$ | $A$ | Ref. |
|---|---|---|---|---|---|---|---|---|
| Nb$_3$Os | 225.26 | 90.87 | 240.31 | 0.40 | 3.35 | 0.32 | 0.48 | This |
|  | 218.78 | - | - | - | - | - | - | [50]$^{Theo}$. |
| Nb$_3$Pt | 199.96 | 85.08 | 223.54 | 0.43 | 3.15 | 0.31 | 0.49 | This |
|  | 201.76 | - | - | - | - | - | - | [50]$^{Theo}$. |
| Nb$_3$Au | 182.62 | 67.74 | 180.86 | 0.37 | 3.87 | 0.33 | 0.41 | This |
|  | 180.88 | - | - | - | - | - | - | [50]$^{Theo}$. |

### *3.3. Debye temperature*

Debye temperature is an important fundamental parameter for solids. It leads to estimates of many important physical parameters such as melting temperature, specific heat, lattice vibration, thermal conductivity, and coefficient of thermal expansion. It also sets the energy scale for electron-phonon interaction in phonon mediated superconductors like the Nb$_3$B phases studied here. The Debye temperature has also related to the vacancy formation energy in metals. Reliable estimates of Debye temperature for different types of materials can be obtained from the elastic moduli [68 – 70]. The average elastic wave velocity, $v_a$, in a crystal is given by

$$v_a = \left[\frac{1}{3}\left(\frac{2}{v_t^3} + \frac{1}{v_l^3}\right)\right]^{-\frac{1}{3}} \quad (4)$$

where, $v_t$ and $v_l$ are the transverse and longitudinal wave velocities, respectively. The transverse velocity $v_t$ is expressed as

$$v_t = \sqrt{\frac{G}{\rho}} \quad (5)$$

where, $\rho$ is the density. The longitudinal wave velocity $v_l$ is obtained from

$$v_l = \sqrt{\frac{B + 4G/3}{\rho}} \quad (6)$$

The Debye temperature, $\Theta_D$, can now be expressed as [71]

$$\Theta_D = \frac{h}{k_B}\left(\frac{3n}{4\pi V_0}\right)^{1/3} v_a \qquad (7)$$

where, $h$ is Planck's constant, $k_B$ is the Boltzmann's constant, $V_0$ is the volume of unit cell and $n$ is the number of atoms within the unit cell.

The calculated Debye temperature $\Theta_D$ of the intermetallics along with the sound velocities $v_l$, $v_t$, and $v_a$ are presented in Table 4.

**Table 4**

Density $\rho$ (in g/cm$^3$), transverse velocity $v_t$ (in ms$^{-1}$), longitudinal velocity $v_l$ (in ms$^{-1}$), average elastic wave $v_a$ velocity (in ms$^{-1}$), and Debye temperature $\Theta_D$ (K), for Nb$_3$B.

| Compounds | $\rho$ | $v_t$ | $v_l$ | $v_a$ | $\Theta_D$ | Ref. |
|---|---|---|---|---|---|---|
| Nb$_3$Os | 10.311 | 2968.7 | 5796.3 | 3066.3 | 226.90 | This |
| | - | - | - | - | 270 | [72]$^{Exp}$ |
| Nb$_3$Pt | 10.228 | 2884.2 | 5535.5 | 2976.6 | 218.98 | This |
| | - | - | - | - | 242 | [72]$^{Exp}$ |
| Nb$_3$Au | 9.958 | 2608.2 | 5235.5 | 2926.3 | 117.97 | This |
| | - | - | - | - | 155 | [72]$^{Exp}$ |

Calculated Debye temperatures show a systematic behavior with the atomic mass. The Debye temperatures, reported in [72], were obtained from the analysis of the specific heat data. Two methods yield somewhat different values of $\Theta_D$ but the trend remains the same.

### 3.4. Electronic density of states and band structure

The calculated total and partial density of states (TDOSs and PDOSs, respectively), as a function of energy, $(E - E_F)$, are presented in Figs. 2(a), (b) and (c), respectively. The vertical broken line denotes the Fermi level, $E_F$. To understand the contribution of each atom to the TDOSs, we have calculated the PDOSs of Nb, Os, Pt and Au atoms in Nb$_3$B. The non-zero values of TDOSs at the Fermi level is evidence that Nb$_3$Os, Nb$_3$Pt, and Nb$_3$Au should exhibit metallic electrical conductivity. At the Fermi energy, the values of TDOSs for Nb$_3$Os, Nb$_3$Pt and Nb$_3$Au are 5.20, 9.77 and 11.50 states per eV per unit cell, respectively. Hence one can say that Nb$_3$Au should be the most conducting among the three. Near the Fermi level the main contribution on TDOSs comes from Nb 4$d$ orbitals. Thus, the electrical conductivity of the compounds is mainly dominated by Nb 4$d$ electronic states. The chemical and mechanical stability of Nb$_3$B are also mainly affected by the properties of Nb 4$d$ bonding electronic states.

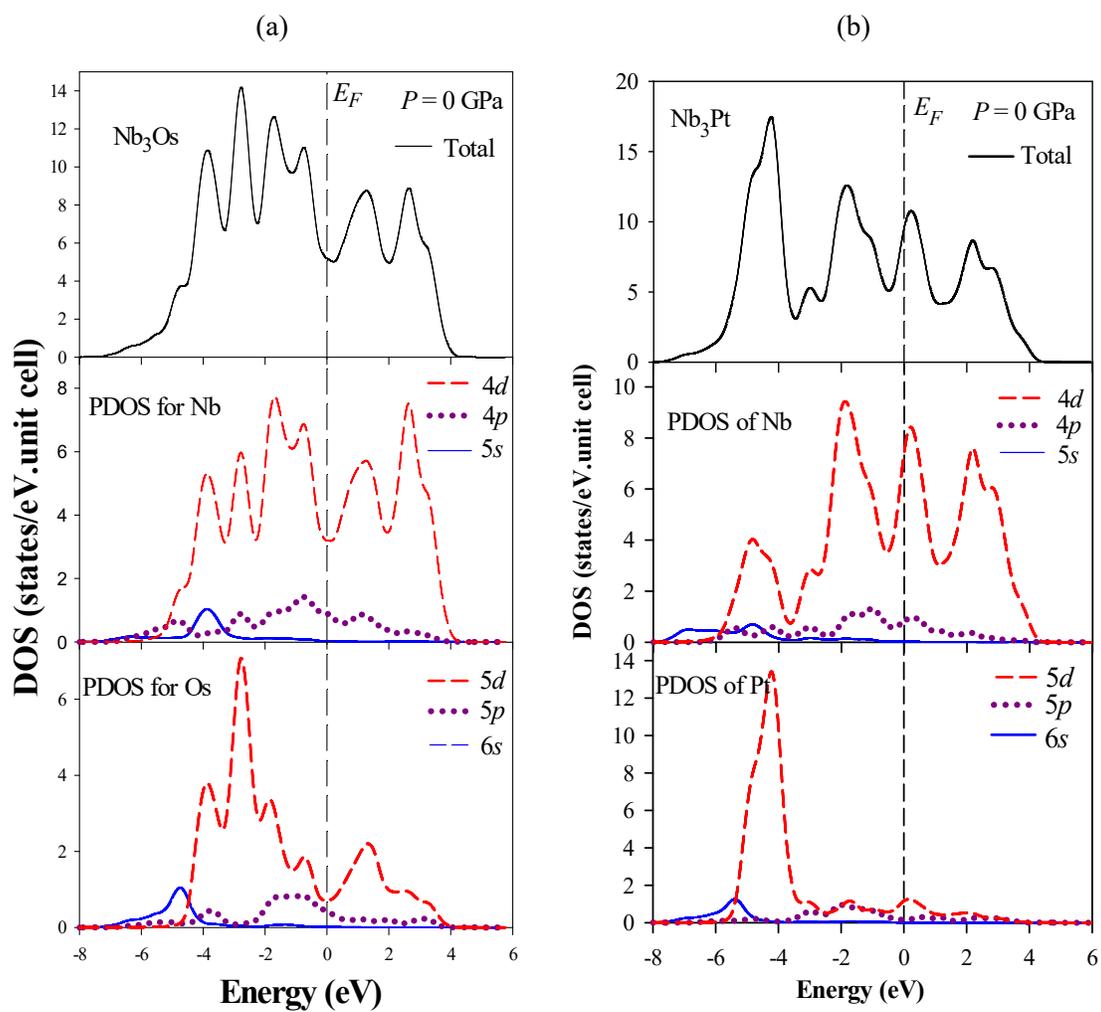

(a)

(b)

(c)

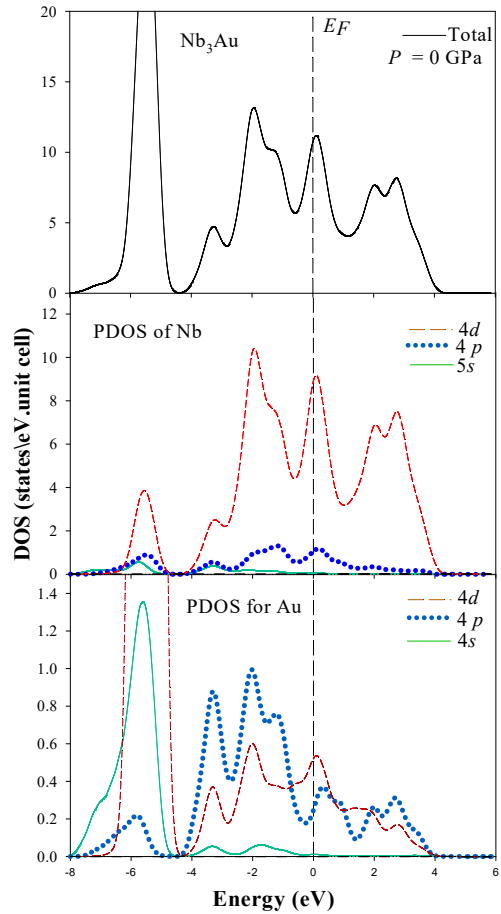

**Figure 2.** Total and partial electronic density of states (DOS) of (a) Nb$_3$Os, (b) Nb$_3$Pt and (c) Nb$_3$Au as a function of energy. The Fermi level is placed at the origin.

From the TDOSs curves of Nb$_3$B, we can see that the valence and conduction bands of Nb$_3$Os is located between -7.34 to 4.18 eV, which is dominated by 4$d$ and 4$p$ states of Nb and with small contribution from 5$p$ and 5$d$ orbitals of Os. For Nb$_3$Pt, the valence band is located between -7.84 and 4.47 eV, which is dominated by Nb 4$d$ orbitals with small contribution from 4$p$ states of Nb and 5$p$, 5$d$ states of Pt. On the other hand, the valence and conduction bands of Nb$_3$Au is dominated by the 4$d$ and 4$p$ states of Nd and some contributions from the 4$d$ orbitals of Au. At low energies there is significant contribution of the 4$s$ electrons of Au atoms to the valence band. The valence band for Nb$_3$Au is splitted. At low energy range from -8.00 to -4.33 eV there is a part of the valence band followed by a region from thereon to 4.01 eV. The valence band strongly overlaps with the conduction band at the Fermi level in Nb$_3$Au. It is also noticed that the $s$ orbitals of Nb and Os/Pt/Au do not contribute to the DOSs at the Fermi level.

Since $Nb_3Os$, $Nb_3Pt$ and $Nb_3Au$ are metallic in nature, the DOSs at Fermi level, $N(E_F)$, is a key parameter for the electronic stability purpose. The phase stability of intermetallic compounds depends on the location of Fermi level and the value of $N(E_F)$ [73-74]. Systems with lower values of $N(E_F)$ are more stable than those with higher values of $N(E_F)$. The total DOSs of $Nb_3Os$, $Nb_3Pt$ and $Nb_3Au$ at the Fermi level are 5.20, 9.77 and 11.50 states per eV per unit cell, respectively. Thus, $Nb_3Os$ is electronically most stable compared to the other two.

Another important feature is the presence of a pseudogap or quasi-gap in the TDOSs in the vicinity of the Fermi level, which is also related to electronic stability [75 - 77]. The pseudogap separates bonding states from nonbonding/antibonding states. For both $Nb_3Os$ the Fermi level lies to the left of the pseudogap i.e., at the nonbonding regions/states, whereas for $Nb_3Pt$, the Fermi level lies to the right of the pseudogap i.e., at the antibonding regions/ states. For $Nb_3Au$, $N(E_F)$ is at a peak of the TDOS.

TDOSs around the Fermi level of $Nb_3Os$ have a number of peaks, which suggest that a dramatic change in physical properties may occur if the Fermi level is shifted by doping, applied stress, etc. It can be seen from TDOSs curves that there are 2 and 4 bonding peaks below the Fermi level for $Nb_3Pt$ and $Nb_3Os$, respectively. For $Nb_3Pt$, strong bonding peak (at -4.27 eV) is formed by the hybridization between Pt $5d$ orbitals and Nb $4s$ orbitals. Similarly, for $Nb_3Os$ a strong peak (at -2.78 eV) is formed due to hybridization between Os $5d$ and Nb $4d$ orbitals. For $Nb_3Au$ the strong peak at the Fermi level is due to the hybridization between $4d$ orbitals of Nb and Au.

We have calculated the electronic band structures of $Nb_3B$ along high symmetry directions ($R$ - $\Gamma$ - $X$ - $M$ - $\Gamma$) in the first Brillouin zone. Figs. 3 show the band structures of these compounds at zero pressure. The Valence Band (VB) and Conduction Band (CB) are considerably overlapped and there is no band gap at the Fermi level.

The total number of bands, as seen from Fig. 3, for $Nb_3Os$, $Nb_3Pt$ and $Nb_3Au$ are 67, 59 and 62, respectively. The bands which cross the Fermi energy are shown (colored) in Figs 3(a), 3(b) and 3(c) with their respective band numbers. The numbers of bands that cross the Fermi energy are 54, 55, 56 and 57 for $Nb_3Os$, 49, 50 and 51 for $Nb_3Pt$ and 49, 50, 51, 52 and 53 for $Nb_3Au$. These bands are both electron- and hole-like, thus resulting in a multiband system dominated by Nb-$d$ electronic states. Specifically hole-like features are observed at symmetry points $M$, $R$, $X$, midway the $X$ - $M$ and $M$ - $\Gamma$ line for $Nb_3Os$ and $R$, $\Gamma$, and midway the $R$ - $\Gamma$ line for $Nb_3Pt$. On the other hand, electron-like features are observed at the symmetry points $\Gamma$, $R$, and $X$ for $Nb_3Os$, $\Gamma$, $X$ and $M$ for $Nb_3Pt$ and $X$ for $Nb_3Au$ . In both $Nb_3Os$, $Nb_3Pt$ and $Nb_3Au$, the lowest energy bands are formed from $s$, $p$, and $d$ states of Os, Pt and Au atoms. The band lying in the range of -3.62 to 4.36 eV for $Nb_3Pt$ mainly arises from $4d$ states of Nb.

Around the Fermi level, there is a tendency of the bands to become more flatter, for both bonding and antibonding orbitals, particularly for the $Nb_3Os$ and $Nb_3Pt$ compounds. The band profiles of $Nb_3Os$ and $Nb_3Pt$ are very similar to $Nb_3Ir$ and $Nb_3Rh$, respectively [78]. The top of the valence bands of $Nb_3Os$, $Nb_3Pt$ and $Nb_3Au$ lie in the energy range from –3.95 eV to Fermi level, –4.83 eV to Fermi level and -4.14 eV to Fermi level, respectively.

A comparison of band structures of $Nb_3Os$, $Nb_3Pt$ and $Nb_3Au$ reveals that the bonding bands are flatter than the antibonding bands around the Fermi level for all the compounds because the lower energy band are more localized than higher energy band. These features of band structure have also been pointed out in literature for several other A-15 systems [79-88]. X-ray photoemission spectra studies of the A-15 type compounds have indicated that the Nb 4*d* and the 5*d* energy bands of these $A_3B$ compounds appear to be more and more separated with increasing atomic number of the *B* atomic species [89-92], as indicated here.

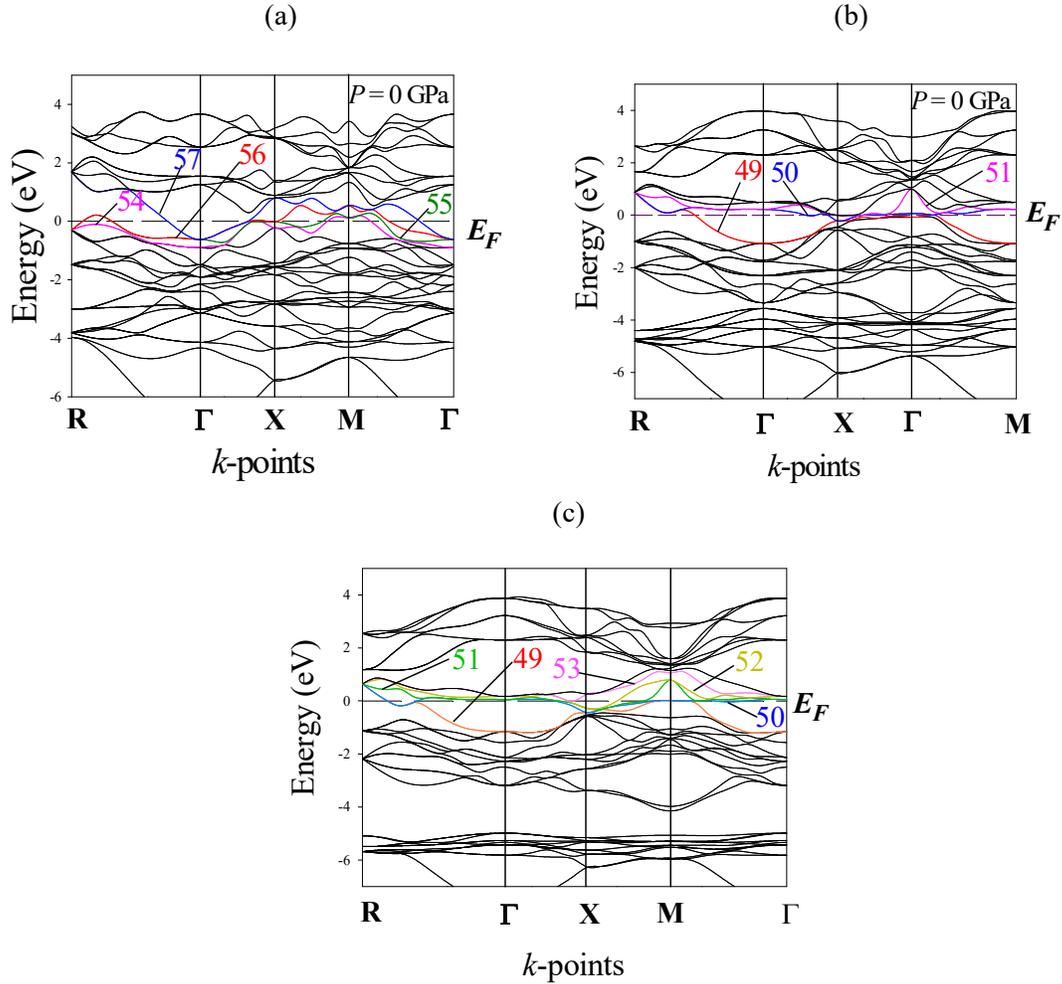

**Figure 3.** Electronic band structure of (a) $Nb_3Os$, (b) $Nb_3Pt$ and (c) $Nb_3Au$ along several high symmetry directions of the Brillouin zone at P = 0 GP.

*3.5. Fermi Surfaces*

Electronic, thermal, optical and some magnetic properties of solids are strongly dependent on the topology of the Fermi surface. 3*D* plots of the Fermi surfaces are shown in Figs. 4, 5 and 6. The Fermi surfaces are constructed from band numbers 54, 55, 56 and 57 for $Nb_3Os$, 49, 50 and 51 for $Nb_3Pt$ and 49, 50, 51, 52 and 53 for $Nb_3Au$. The Fermi surfaces of $Nb_3Os$, $Nb_3Pt$ and $Nb_3Au$ contain both electron- and hole-like sheets. For band 54 of $Nb_3Os$, hole-like sheet is seen around the *M*-point. The Fermi surface topology for band 55 of $Nb_3Os$ is quite complex. There is a hole-like sheet around *M*-point and electron-like sheets around *X*- and *R*-points. For band 56 of $Nb_3Os$, a tiny electron-like sheet appears around the point *R* and electron-like sheet is seen around the central *Γ*-point. For band 57 an electron-like sheet forms around *Γ*-point. For band 49 of $Nb_3Pt$, a hole like curvature is found around the symmetry point R. These Fermi sheets are separated from each other. For both the bands 50 and 51 of $Nb_3Pt$, electron-like sheets are formed around the *X*-point. For band 51, this sheet is small. For band 49 of $Nb_3Au$, hole-like sheets appear around R- and M-points. For band 50 of $Nb_3Au$, two separate hole-like sheets present, among them one is around *Γ*-point. For band 51and 52 of $Nb_3Au$, two electron-like Fermi sheets form around *X*-point. For band 53 of $Nb_3Au$, few very tiny Fermi surfaces appear.

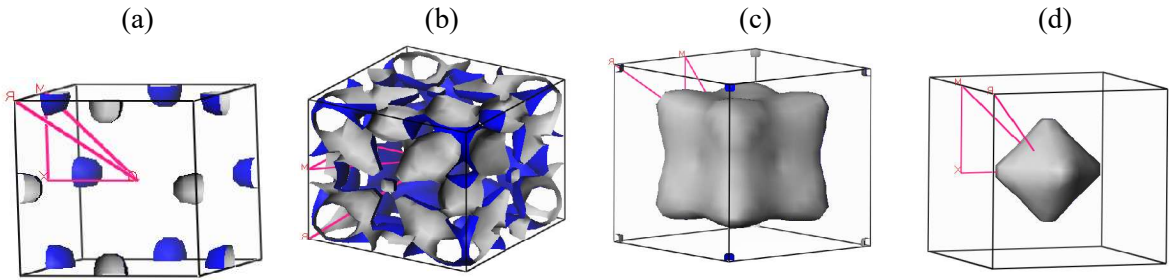

**Figure 4.** Fermi surface for bands (a) 54, (b) 55, (c) 56 and (d) 57 of $Nb_3Os$, respectively.

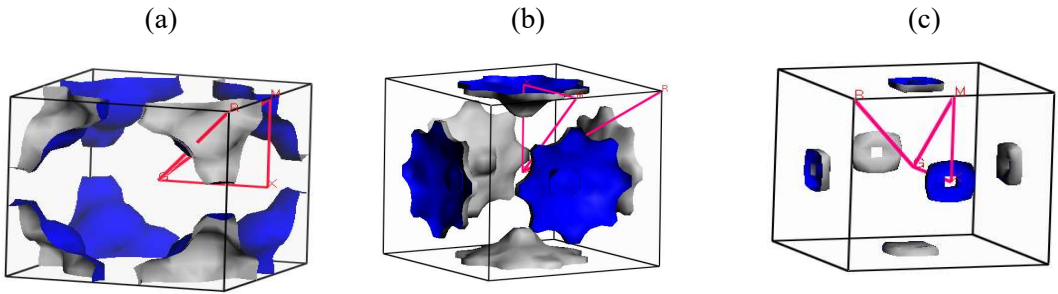

**Figure 5.** Fermi surface for bands (a) 49, (b) 50 and (c) 51 of $Nb_3Pt$, respectively.

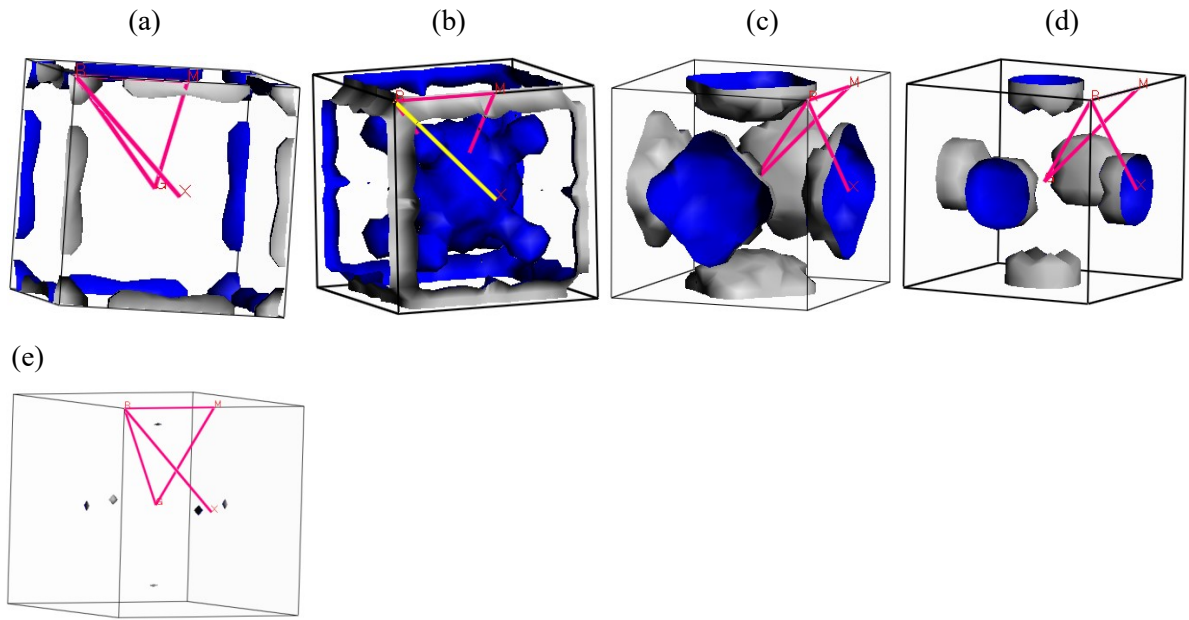

**Figure 6.** Fermi surface for bands (a) 49, (b) 50, (c) 51, (d) 52 and (e) 53 of $Nb_3Au$, respectively.

*3.6. Electronic charge Density*

In order to investigate the nature of electronic bonding in $Nb_3Os$, $Nb_3Pt$ and $Nb_3Au$, we have studied the electronic charge density distribution in the (110) plane (Figs. 7). The charge density distribution map shows that there is an ionic bonding between Nb-Os, Nb-Pt and Nb-Au atoms and covalent bonding between Nb-Nb atoms for $Nb_3Os$, $Nb_3Pt$ and $Nb_3Au$. These agree with the Mulliken bonding population analysis (Section 3.7).

The color scale on the right side of charge density maps shows the total electron density. The blue color shows the high charge (electron) density and red color shows low charge (electron) density. So, Os atoms have high electron density compared to Nb atoms for $Nb_3Os$, this also agrees with the Mulliken charge analysis. On the other hand, Nb atoms have greater charge density than Pt atoms for $Nb_3Pt$. This is may be due to the *d* orbital of Pt atoms, which is confined in space and whose density of states are quite low. This is supported by Mulliken bonding population analysis. From the charge density map of $Nb_3Au$, Nb atoms have high charge density compared to Au atoms.

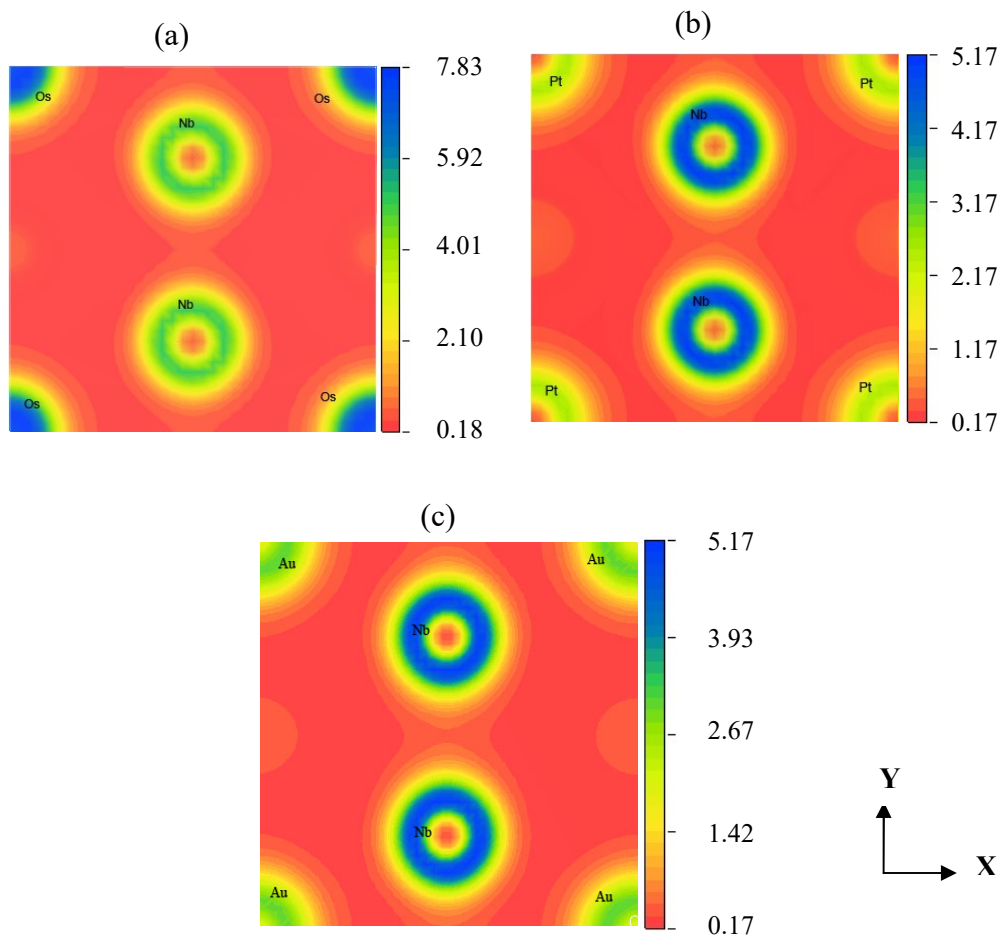

**Figure 7.** The electronic charge density map for (a) Nb$_3$Os, (b) Nb$_3$Pt and (c) Nb$_3$Au in the (110).

*3.7. Bond Population Analysis*

To explore the bonding nature in greater depth, the Mulliken bond populations [93] are studied. The results of this analysis are given in Table 5. The atomic charge density of Nb and Os in Nb$_3$Os are 0.15 and -0.46 electron, respectively. Both deviates from the normal value expected in a purely ionic state (Nb: +5, +3 and Os: +4). This deviation again partially reflects the covalent bonding character between Nb species that forms by the hybridization between *d* states of Nb, and strong ionicity, which again is in good agreement with the charge density analysis. Similarly the atomic charge of Nb and Pt in Nb$_3$Pt are 0.21 and -0.62 electron, respectively. Both these values deviated from the normal value expected for a purely ionic state (Nb: +5, +3 and Pt: +4, +2). These deviations again partially reflect the covalent character of the bond between Nb atoms (*d-d* hybridized covalent bonding between Nb) and strong iconicity, which is in good agreement with the charge density mapping shown in Fig. 7. On the other hand, the atomic charge of Nb and Au in Nb$_3$Au are 0.20 and -0.60 electron, respectively. Both these values deviated from the normal value expected for a purely ionic state (Nb: +5, +3 and Au: +1, +3). These deviations partially reflect the covalent character of the bond between Nb atoms (*d-d* hybridized covalent bonding between Nb) and strong iconicity, which is also in good agreement with the charge density mapping. The charge of Nb in Nb$_3$Os is 0.06 and 0.05 electrons smaller than that of Nb in Nb$_3$Pt and Nb$_3$Au, respectively. This difference reflects that Nb in Nb$_3$Pt release more electrons into the conduction band than Nb in Nb$_3$Os and Nb$_3$Au. The atomic charge of Os, Pt and Au are -0.46, -0.62 and -0.60 electron, respectively. The atomic charge of Os is 0.16 and 0.14 electrons larger than that of Pt and Au respectively, which is also indicative of the release of 0.16 more electrons into the conduction band by Os. It is observed that the band spilling parameters are quite low for both the compounds. This results in relatively lower TDOS at the Fermi level for these two compounds as revealed by our electronic band structure calculations.

For all Nb$_3$Os, Nb$_3$Pt and Nb$_3$Au, electrons are transferred from Nb to Os, Pt and Au, respectively, suggesting that an ionic contribution to the bonding is present. Thus we can say that Nb-Os, Nb-Pt and Nb-Au bonding are largely ionic in Nb$_3$Os, Nb$_3$Pt and Nb$_3$Au. This is in good agreement with the charge density mapping. The degree of covalency and/or ionicity may be obtained from the effective valence. This is defined as the difference between the formal ionic charge and the Mulliken charge on the cation species [94]. The zero value of effective charge indicates a perfectly ionic bond, while values greater than zero indicate an increasing level of covalency. The effective valence for Nb in Nb$_3$Os, Nb$_3$Pt and Nb$_3$Au is +2.85, +2.79 and +2.80 electron, respectively. All these indicate that both ionic and covalent bonds are present in Nb$_3$Os, Nb$_3$Pt and Nb$_3$Au.

Because of the strong basis set dependence of Mulliken Population Analysis (MPA), sometimes MPA gives results in contradiction to chemical intuition. For this reason we have also determined Hirshfeld charge using Hirshfeld Population Analysis (HPA). This has practically no basis set dependence and can provide with a more physically meaningful result compared to MPA. Table 5 shows the comparison between Mulliken and Hirshfeld charge. Here we note that the magnitude of Hirshfeld charge is in general smaller than that of Mulliken charge. The gross features of the bonding nature remain unchanged. The level of covalency and iconicity is lower in the HPA. This should imply a greater contribution of metallic bonding in Nb$_3$Os, Nb$_3$Pt and Nb$_3$Au which is supported by the analysis of the elastic constant data.

**Table 5**
Charge spilling parameter (%), orbital charges (electron), atomic Mulliken charges (electron), effective valence (electron) and Hirshfeld charge (electron) in Nb$_3$Os, Nb$_3$Pt and Nb$_3$Au.

| Compounds | Species | Charge spilling | $s$ | $p$ | $d$ | Total | Mulliken charge | Effective valence | Hirshfeld charge |
|---|---|---|---|---|---|---|---|---|---|
| Nb$_3$Os | Nb |  | 2.22 | 6.65 | 3.97 | 12.85 | 0.15 | +2.85 | 0.09 |
|  | Os | 0.06 | 2.71 | 7.14 | 6.61 | 16.46 | -0.46 |  | -0.28 |
| Nb$_3$Pt | Nb |  | 2.30 | 6.52 | 3.97 | 12.79 | 0.21 | +2.79 | 0.02 |
|  | Pt | 0.10 | 0.83 | 1.13 | 8.66 | 10.62 | -0.62 |  | -0.07 |
| Nb$_3$Au | Nb |  | 2.20 | 6.59 | 4.01 | 12.80 | 0.20 | +2.80 | -0.01 |
|  | Au | 0.05 | 0.92 | 1.28 | 9.39 | 11.60 | -0.60 |  | 0.04 |

*3.8. Vickers hardness of Nb₃B*

The calculated bond hardness and the Vickers hardness of the compounds understudy are presented in the table shown below.

**Table 6**

The calculated Mulliken bond overlap population of $\mu$-type bond $P^\mu$, bond length $d^\mu$(Å), metallic population $P^{\mu'}$, total number of bond $N^\mu$, bond volume $v_b^\mu$(Å³), hardness of $\mu$-type bond $H_v^\mu$(GPa) and Vickers hardness of the compound, $H_v$(GPa) of Nb₃Os, Nb₃Pt and Nb₃Au.

| Compounds | Bond | $d^\mu$ | $P^\mu$ | $P^{\mu'}$ | $N^\mu$ | $v_b^\mu$ | $H_v^\mu$ | $H_v$ |
|---|---|---|---|---|---|---|---|---|
| Nb₃Os | Nb-Nb | 2.58 | 0.91 | 0.072 | 15 | 6.859 | 24.80 | 9.764 |
|  | Nb-Pt | 2.88 | 0.52 | 0.072 |  | 9.540 | 7.625 |  |
| Nb₃Pt | Nb-Nb | 2.59 | 0.68 | 0.024 | 15 | 7.043 | 18.932 | 7.584 |
|  | Nb-Os | 2.90 | 0.39 | 0.024 |  | 9.886 | 5.958 |  |
| Nb₃Au | Nb-Nb | 2.62 | 0.64 | 0.031 | 15 | 7.284 | 16.449 | 5.710 |
|  | Nb-Au | 2.93 | 0.31 | 0.031 |  | 10.188 | 4.336 |  |

From Table 6 it is seen that Nb₃Os is the hardest among the compounds studied here. The hardness of Nb₃Os is comparable to many technologically important MAX phase nanolaminates and to some ternary borides and borocarbides [95-98]. Nb₃Au has the lowest value of Vickers hardness. Comparatively high hardness of Nb₃Os follows from the presence of strong covalent bonds in this compound. The electronic band structure calculation also indicates that, among the three compounds studied here, Nb₃Os is structurally the most stable. Comparatively higher value of Debye temperature and elastic moduli of Nb₃Os also support the hardness analysis presented in Table 6.

## 4. Conclusions

First principles calculations based on the DFT have been used to investigate a number of physical properties of niobium based intermetallic Nb₃Os, Nb₃Pt and Nb₃Au compounds. The equilibrium lattice parameters at zero pressure of these compounds are in very good agreement with both previously reported theoretical and experimental values. The calculated elastic parameters allow us to conclude that all these compounds are mechanically stable. Young's modulus of Nb₃Os is substantially larger than Nb₃Pt and Nb₃Au which indicates that Nb₃Os is much stiffer than Nb₃Pt and Nb₃Au. Poisson's ratio, Cauchy pressure, and Pugh's ratio indicate that Nb₃Os, Nb₃Au and Nb₃Au should be ductile metallic material. The calculated shear anisotropy factors imply that all the compounds are elastically anisotropic. From $B/C_{44}$ values, we can conclude that Nb₃Au is expected to have good lubricating properties and high degree of machinability compared to Nb₃Os and Nb$_3$Pt. The Debye temperatures of the intermetallics are low. The calculated values of $\Theta_D$ using the estimated elastic moduli agree reasonably well with experimental values [72]. Since these materials are predicted to be metallic, conduction of heat is expected to be done predominantly by electrons at low temperatures. All three compounds are superconducting at low temperatures. Nb$_3$Au has the highest superconducting transition temperature (11 K), followed by Nb$_3$Pt (10 K) and Nb$_3$Os (0.94 K) [28]. It is interesting to note that Nb$_3$Au has a very low Debye temperature (therefore the characteristic phonon energy) but the DOS at the Fermi level is quite high. This implies that the electron phonon coupling parameter, $\lambda$ ($\lambda = N(E_F)V_{e\text{-ph}}$, where $V_{e\text{-ph}}$ is the electron-phonon interaction energy) should be quite high in this compound [99]. Nb$_3$Os, on the other hand has a very low $T_c$ even if the Debye temperature is the highest among the compounds studied here. This indicates that the electron phonon coupling is very weak for Nb$_3$Os. Considering the high value of $T_c$ for Nb$_3$Pt, a reasonably high value of $\lambda$ can be expected. The electronic energy band structure and DOS analyses reveal that all the materials are metallic in nature, where the conductivity is expected to be dominated by the Nb-4$d$ electronic orbitals. Considering the value of $N(E_F)$, electrical conductivity of Nb$_3$Au should be the highest followed by Nb$_3$Pt and Nb$_3$Os. DOS profiles indicate that Nb$_3$Os is electronically the most stable. Fermi surfaces of the intermetallics exhibit a complex combination of electron and hole like sheets. Result of Mullikan population analysis and charge density mapping suggest that Nb$_3$Os, Nb$_3$Pt and Nb$_3$Au have both ionic and covalent bondings in addition to metallic bonding. For all the compounds electrons are transferred from Nb to Os, Pt or Au. Maximum charge density is found around Os for Nb$_3$Os, Nb for Nb$_3$Pt and Nb$_3$Au atoms, respectively. Mulliken population analysis also shows that Nb-Nb bonding is stronger than Nb-Os/Nb-Pt/Nb-Au bonding in Nb$_3$Os, Nb$_3$Pt and Nb$_3$Au, respectively. The Vickers hardness is quite high for Nb$_3$Os followed by Nb$_3$Pt and Nb$_3$Au.

Very recently Li et al. [100] have reported some DFT based results on the structural, elastic, electronic and charge distribution properties of Nb$_3$Ir and Nb$_3$Pt compounds. The gross features of their results regarding Nb$_3$Pt show fair agreement with those presented here, where applicable.

To conclude, this study presents a detailed DFT based *ab-initio* study of elastic, electronic, Fermi surface features and bonding properties of technologically important intermetallic compounds Nb$_3$Os, Nb$_3$Pt and Nb$_3$Au for the first time. We hope that this work will inspire other research groups to study these interesting materials further in the near future.

# References

1. Hume-Rothery, W.: J. Ins. Metals **35** 209 (1925)
2. J. Liang, D. Fan, P. Jiang, H. Liu, and W. Zhao, Intermetallics **87** 27 (2017)
3. T. An and F. Qin, Journal of Electronic Packaging **138** DOI: 10.1115/1.4032349 (2015)
4. H. Lu, N. Zou, X. Zhao, J. Shen, X. Lu, Y. He, Intermetallics **88** 91 (2017)
5. Y. Terada, K. Ohkubo, S. Miura, J.M. Sanchez, T. Mohri, Journal of Alloys and Compounds **354** 202 (2003)
6. M. Rajagopalan, M. Sundareswari, Journal of Alloys and Compounds **379** 8 (2004)
7. Y. Terada, Platinum Metals Rev. **52** 208 (2008)
8. L. Mohammedi, B. Daoudi, A. Boukraa, Computational Condensed Matter **2** 11 (2015)

9. J. Magnien, G. Khatibi, M. Lederer, H. Ipser, DOI: 10.1016/j.msea.2016.07.060
10. W. Zhong, F. Qin, T. An, T. Wang, DOI: 10.1109/ICEPT.2010.5582669
11. K.S. Kim, S.H. Hun, K. Suganuma, Journal of Alloys and Compounds **352** 226 (2003)
12. M.H.F. Sluiter, Calphad **30** 357 (2006)
13. K. Tachikawa, Fusion Eng. Des. **81** 2401 (2006)
14. A. Godeke, B. Haken, et al., Supercond. Sci. Technol. **19** R100 (2006)
15. R. Boscencu, M. Ilie, R. Socoteanu, Int. J. Mol. Sci. **12** 5552 (2011)
16. H. Ohno, T. Shinoda, Y. Oya-Seimiya, J. Japan Inst. Met. **68** 769 (2004)
17. H. Kumakura, H. Kitaguchi, A. Matsumoto et al., Supercond Sci.Technol. **18** 147 (2005)
18. A. Godeke, M. C. Jewell, C. M. Fischer et al., J. Appl. Phys. **97** 1 (2005)
19. P.J. Lee, D.C. Larbalestier, IEEE Trans. Appl. Supercond. **15** 3474 (2005)
20. C. V. Renaud, T. Wong, L.R. Motowidlo, IEEE Trans. Appl. Supercond. **15** 3418 (2005)
21. S. Haindl, M. Eisterer, R. Muller et al., IEEE Trans. Appl. Supercond. **15** 3414 (2005)
22. T. Takeuchi, M. Kosuge, N. Banno et al., Supercond. Sci. Technol. **18** 985 (2005)
23. A.V. Skripov, L.S. Voyevodina, R. Hempelmann, Phys. Rev. B **73** 1 (2006)
24. C.D. Hawes, P.J. Lee, D.C. Larbalestier, Supercond. Sci. Technol. **19** S27 (2006)
25. S.M. Deambrosis, G. Keppel et al., Physica C **441** 108 (2006)
26. A. Godeke, Supercond. Sci. Technol. **19** R68 (2006)
27. G.R. Stewart, Physica C: Superconductivity and its Applications **514** 28 (2015)
28. D. Dew-Hughes, Cryogenics **15** 435 1975
29. Y. Ding, S. Deng, Y. Zhao, Journal of Modern Transportation **22** 183 (2014)
30. C. Paduani, Brazilian Journal of Physics **37** 1073 (2007)
31. B.M. Klein, L.L. Boyer, D.A. Papconstantopoulos, Phys. Rev. Lett. **42** 530 (1979)
32. E.Z. Kurmaev, F. Werfel, O. Brümmer, R. Flükiger, Solid State Commun. **21** 39 (1977)
33. E.Z. Kurmaev, V.P. Belash, R. Flukiger, A. Junod, Solid State Commun. **16** 1139 (1975)
34. R.A. Pollak, C.C. Tsuei, R.W. Johnson, Solid State Commun. **23** 879 (1977)
35. A. Junod, T. Jarlborg, Muller, J Phys. Rev. B **27** 1568 (1983)
36. Materials studio CASTEP manual © Accelrys 2010
    http://www.tcm.phy.cam.ac.uk/castep/documentation/WebHelp/CASTEP.html
37. M.D. Segall, P.J.D. Lindan, M.J. Probert, C.J. Pickard, P.J. Hasnip, S.J. Clark, M.C. Payne, J. Phys.: Condens. Matter **14** 2717 (2002)
38. M.C. Payne, M.P. Teter, D.C. Allan, T.A. Arias, J.D. Joannopoulos, Rev. Mod. Phys. **64** 1045 (1992)
39. D. Vanderbilt, Phys. Rev. B **41** 7892 (1990)
40. J.P. Perdew, K. Burke, M. Ernzerhof, Phys. Rev. Lett. **77** 3865 (1996)
41. H.J. Monkhorst, J.D. Pack, Phys. Rev. B **13** 5188 (1976)
42. T.H. Fischer, J. Almlof, J. Phys. Chem. **96** 9768 (1992)
43. F.D. Murnaghan, Finite Deformation of an Elastic Solid. John Wiley, New York. 1951
44. D. Sanchez-Portal, E. Artacho, J.M. Soler, Solid State Commun. **95** 685 (1995)
45. M.D. Segall, R. Shah, C.J. Pickard, M.C. Payne, Phys. Rev. B **54** 16317 (1996)
46. S. Geller ibid. **9** 885 (1956)
47. S.V. Reddy, S.V. Suryanarayana, Journal of Materials Science Letters **3** 763 (1984)

48. P.A. Beck (Ed.), Electronic Structure and Alloy Chemistry of the Transition Elements, Interscience Publishers, New York, 1963
49. M.V. Nevit, in: J. H. Westbrook (Ed.), Intermetallics Compounds, R. E. Krieger Publishing Co., Huntington NY, 1977
50. C. Paduani, Solid State commun. **144** 352 (2007)
51. M. Mattesini, R. Ahuja, B. Johansson, Phys. Rev. B **68** 184108 (2003)
52. A. Sari, G. Merad, H. Si Abdelkader, Computational Materials Science **96** 348 (2015)
53. M.E. Fine, L.D. Brown, H.L. Marcus, Scr. Metall. **18** 951 (1984)
54. R.L. Fleischer, in Intermetallic Compounds, Edited by O Izumi, Japan Inst. Met., Sendai (1991a), p. 157
55. J. Haines, J.M. Leger, G. Bocquillon, Annu. Rev. Mater. Res. **31** 1 (2001)
56. Q.M. Hu, R. Yang, Curr. Opin. Solid St. Mater. Sci. **10** 19 (2006)
57. B.Y. Tang, W.Y. Yu, X.Q. Zeng, W.J. Ding, M.F. Gray, Mater. Sci. Eng. **A** *489* 444 (2008)
58. S.F. Pugh, Philos. Mag. **45** 43 (1954)
59. V.V. Bannikov, I.R. Shein, A.L. Ivanovskii, Physica B **405** 4615 (2010)
60. W. Feng, S. Cui, Can. J. Phys. **92**: 1652 dx.doi.org/10.1139/cjp-2013-0746 (2014)
61. Z. Sun, D. Music, R. Ahuja, J.M. Schneider, Phys. Rev. B **71** 193402 (2005)
62. L. Vitos, P.A. Korzhavyi, B. Johansson, Nature Mater. **2** 25 (2003)
63. R.C. Lincoln, K.M. Koliwad, P.B. Ghate, The phys. Rev. **157** 463 (1967)
64. K.J. Puttlitz, K.A. Stalter, *Handbook of Lead-Free Solder Technology for Microelectronic Assemblies (*New York: Springer) p. 98 (2005)
65. M.A. Ali, A.K.M.A. Islam, M.S. Ali, J. Sci. Res. **4** 1 (2012)
66. M.J. Phasha, P.E. Ngoepe, H.R. Chauke, D.G. Pettifor, D. Nguyen-Mann, Intermetallics **18** 2083 (2010)
67. M. Sundareswari, S. Ramasubramanian, M. Rajagopalan, Solid State Commun. **150** 2057 (2010)
68. M.A. Ali, M.A. Hadi, M. M. Hossain, S.H. Naqib, A.K.M.A. Islam, Phys. Status Solidi (b) DOI: 10.1002/pssb.201700010 (2017)
69. M.A. Hadi, M.S. Ali, S.H. Naqib, A.K.M.A. Islam, Chin. Phys. B **26** 037103 (2017)
70. M.A. Hadi, M.T. Nasir, M. Roknuzzaman, M.A. Rayhan, S.H. Naqib, A.K.M.A. Islam, Phys. Status Solidi (b) **253** 2020 (2016)
71. M.A. Hadi, M. Roknuzzaman, A. Chroneos, S.H. Naqib, A.K.M.A. Islam, V. Vovk, K. Ostrikov, Comp. Mat. Sci. **137** 318 (2017)
72. S.V. Reddy, S.V. Suryanarayana, Journal of Mat. Sci. Lett. **5** 436 (1986)
73. J.H. Xu, T. Oguchi, A.J. Freeman, Phys. Rev. B **36** 4186 (1987)
74. T. Hong, T.J. Watson-Yang, A.J. Freeman, T. Oguchi, J.H. Xu, Phys. Rev. B **41** 12462 (1990)
75. C.D. Gelatt, Jr. A.R. Williams, V.L. Mourzzi, Phy. Rev. B **27** 2005 (1983)
76. A. Pasturel, C. Colinet, P. Hicter, Physica B **132** 177 (1985)
77. I. Galanakis, P. Mavropoulous, J. Phys.: Condens. Matter **19** 315213 (2007)
78. C. Paduani, Physica B **393** 105 (2007)

79. G. Arbman, T. Jarlborg, Solid State Commun. **26** 857 (1978)
80. T. Jarlborg, A. Junod, M. Peter, Phys. Rev. B **27** 1558 (1983)
81. K.M. Ho, M.L. Cohen, W.E. Pickett, Phys. Rev. Lett. **41** 815 (1978)
82. A.T. Van Kessel, H.W. Myron, F.M. Mueller, Phys. Rev. Lett. **41** 181 (1978)
83. W.E. Pickett, K.M. Ho, M.L. Cohen, Phys. Rev. **19** 1734 (1979)
84. L.F. Mattheiss, W. Weber, Phys. Rev. B **25** 2243 (1982)
85. B. Sadigh, V. Ozolins, Phys. Rev. B **57** 2793 (1998)
86. B.M. Klein, L.L. Boyer, D.A. Papconstantopoulos, Phys. Rev. Lett. **42** (8) 530 (1979)
87. C. Paduani, Solid State commun. **144** 352 (2007)
88. C. Paduani, Physica B **393** 105 (2007)
89. E.Z. Kurmaev, F. Werfel, O. Brümmer, R. Flükiger, Solid State Commun. **21** 39 (1977)
90. E.Z. Kurmaev, V.P. Belash, R. Flukiger, A. Junod, Solid State Commun. **16** 1139 (1975)
91. R.A. Pollak, C.C. Tsuei, R.W. Johnson, Solid State Commun. **23** 879 (1977)
92. A. Junod, T. Jarlborg, J. Muller, Phys. Rev. B **27** 1568 (1983)
93. R.S. Mulliken, J. Chem. Phys. **23** 1833 (1955)
94. M.D. Segall, R. Shah, C.J. Pickard, M.C. Payne, Phys. Rev. B **54** 16317 (1996)
95. M.A. Hadi, S.-R.G. Christopoulos, S.H. Naqib, A. Chroneos, M.E. Fitzpatrick, A.K.M.A. Islam, Journal of Alloys and Compounds **748** 804 (2018)
96. M.A. Hadi, S.H. Naqib, S.-R.G. Christopoulos, A. Chroneos, A.K.M.A. Islam, Journal of Alloys and Compounds **724** 1167 (2018)
97. F. Parvin, S.H. Naqib, Chin. Phys. B **26** 106201 (2017)
98. P. Barua, M.M. Hossain, M.A. Ali, M.M. Uddin, S.H. Naqib, A.K.M.A. Islam, arXiv:1805.03392
99. M.A. Hadi, M.A. Alam, M. Roknuzzaman, M.T. Nasir, A.K.M.A. Islam, S.H. Naqib, Chin. Phys. B **24** 117401 (2015)
100. Xianfeng Li, Dong Chen, Yi Wu, Mingliang Wang, Naiheng Ma, Haowei Wang, AIP Advances **7** 065012 (2017)